\documentclass[12pt]{article}

\usepackage[numbers,sort&compress]{natbib}
\usepackage{amsmath}
\usepackage{amssymb}
\usepackage{hyperref}

\usepackage{sectsty}
\sectionfont{\large}
\subsectionfont{\normalsize}

\newcommand{\ba}{\begin{aligned}}
\newcommand{\ea}{\end{aligned}}
\newcommand{\beq}{\begin{equation}}
\newcommand{\eeq}{\end{equation}}
\def\ben{\begin{equation*}}
\def\een{\end{equation*}}
\newcommand{\beqs}{\begin{eqnarray}}
\newcommand{\eeqs}{\end{eqnarray}}

\newcommand{\sdot}{\hspace{-3pt}\cdot\hspace{-3pt}}

\def\be{\begin{equation}}
\def\ee{\end{equation}}
\def\bea{\begin{eqnarray}}
\def\eea{\end{eqnarray}}
\def\bsp{\be\begin{split}}

\def\a{\alpha}

\def\d{\delta}
\def\e{\epsilon}

\def\s{\sigma}

\makeatletter

\newcommand{\Rmnum}[1]{\expandafter\@slowromancap\romannumeral #1@}
\makeatother

\renewcommand{\title}[1]{\vbox{\center\LARGE{#1}}\vspace{5mm}}
\renewcommand{\author}[1]{\vbox{\center\large{#1}}\vspace{5mm}}

\everymath{\displaystyle}

\begin{document}

\begin{titlepage}
\vspace{10pt} \hfill {} \vspace{20mm}
\begin{center}

{\Large \bf Scattering equations, generating functions and all massless five point tree amplitudes}

\vspace{45pt}

{
\textbf{Chrysostomos Kalousios},$^a$
\footnote[1]{\href{mailto:ckalousi@ift.unesp.br}{\tt{ckalousi@ift.unesp.br}}}
}
\\[15mm]

{\it\ ${}^a\,$ICTP South American Institute for Fundamental Research\\
Instituto de F\'\i sica Te\'orica, UNESP-Universidade Estadual Paulista\\
R. Dr. Bento T. Ferraz 271 - Bl. II, 01140-070, S\~ao Paulo, SP, Brasil}\\
\vspace{10pt}

\vspace{20pt}

\end{center}

\vspace{40pt}

\centerline{{\bf{Abstract}}}
\vspace*{5mm}
\noindent
We argue that one does not need to know the explicit solutions of the scattering equations in order to evaluate a given amplitude.  We consider the most general quantity consistent with $SL(2,\mathbb{C})$ invariance that can appear in an amplitude that admits a scattering equation description.  This quantity depends on all cross ratios that can be formed from $n$ points and we evaluate it for the first non-trivial case of $n=5$.  The combinatorial nature of the problem is captured through the construction of an appropriate generating function that depends on five variables.

\vspace{15pt}
\end{titlepage}

\newpage

\section{Introduction}
In \cite{Cachazo:2013gna} it was argued that the tree level S-matrix of massless theories can be captured by the so-called scattering equations (to be defined later) that connect the space of kinematic invariants of $n$ particles in arbitrary spacetime dimensions to the positions of $n$ points on a sphere.  Prior to \citep{Cachazo:2013gna} the scattering equations had appeared in the literature in different contexts in \cite{Fairlie, Roberts, Fairlie:2008dg, Gross:1987ar, Witten:2004cp, Caputa:2011zk, Caputa:2012pi, Makeenko:2011dm, Cachazo:2012uq}. 

After the initial conception and application of the scattering equations to Yang-Mills and gravity \cite{Cachazo:2013hca}, an increasing number of theories, whose tree level amplitudes can be expressed in terms of the scattering equations, has been found \cite{Cachazo:2013iea, Dolan:2013isa, Naculich:2014naa, Cachazo:2014nsa, Cachazo:2014xea, Weinzierl:2014ava, Naculich:2015zha} some of them generalizing to massive cases, with the promise that this is not the complete list.  It is still not known what kind of theories admit a representation in terms of scattering equations.  

The formalism was proven for Yang-Mills in \cite{Dolan:2013isa}, where the authors showed that it reproduces the BCFW \cite{Britto:2005fq} recursion relations.  A polynomial form of the scattering equations that greatly facilitates computations was also presented in \cite{Dolan:2014ega}.  Extensions at loop level include \cite{Adamo:2013tsa, Casali:2014hfa, Adamo:2015hoa} and connection to twistor-string-like models can be found in \cite{Mason:2013sva, Berkovits:2013xba, Geyer:2014fka}.

One of the questions is how to use the formalism in order to get explicit answers for the amplitudes.  Some attempts have appeared in the past and involved special solutions of the equations  associated to particular polymonials \cite{Cachazo:2013iea, Dolan:2014ega, Kalousios:2013eca, Weinzierl:2014vwa, Lam:2014tga}, that in some of the cases allowed for the explicit construction of the amplitude \cite{Cachazo:2013iea, Kalousios:2013eca}.  Besides all the  efforts there is no known general solution of the scattering equations.  

Fortunately, one does not need to know the explicit solutions of the equations in order to evaluate the amplitude.  The amplitude is always given as sums of all possible solutions of the scattering equations, which are polynomial in nature, and one can then use the well known in mathematics formulas of Vieta, that associate the sums of roots of polynomials to the coefficients of these polynomials.  This can be the first step, but still the answer is complicated to evaluate and write it down.

In this work we attempt to organize the expressions of tree level amplitudes.  We first identify a fundamental quantity, whose general form depends on products of cross ratios.  We claim that all amplitudes can be written as linear combinations of this quantity.  In the case of $n=5$, we explicitly evaluate this fundamental quantity by constructing a generating function that captures the combinatorics of the problem.  We then give specific examples for the case of Yang-Mills.

\section{Scattering equation formalism}
The fundamental ingredient that allows the formulation of S-matrices in arbitrary dimensions is the scattering equations and are defined as
\be \label{scattering equations}
f_a = \sum_{b \neq a}^n \frac{k_{ab}}{\s_{ab}},
\ee
where $k_a$ is the momentum of the $a^{\rm th}$ particle.  In the above we have used the short notation $k_{ab}=k_a \sdot k_b$ and $\s_{ab}=\s_a-\s_b$.  Not all of the $n$ equations in \eqref{scattering equations} are independent, but instead they satisfy three constraints
\be 
\sum_{a=1}^n f_a = \sum_{a=1}^n \s_a f_a = \sum_{a=1}^n \s_a^2 f_a = 0.
\ee
This is due to the $SL(2,\mathbb{C})$ invariance of \eqref{scattering equations}, which is a direct consequence of total momentum conservation and the on-shell condition of the external particles.  This allows us to fix three of the $\s_i$s to arbitrary values.  The number of solutions of \eqref{scattering equations} is known \cite{Cachazo:2013gna} to be $(n-3)!$ and in general they can be complex.

The S-matrices of the theories that admit a scattering equation description have the general form
\be \label{Mn=}
\mathcal{M}_n = \int \frac{{\rm d}^n \s}{{\rm vol}\, SL(2,\mathbb{C})}
\s_{ij}\s_{jk}\s_{ki}\hspace{-6pt} \prod_{a \neq i,j,k}\hspace{-6pt} \d(f_a) \hspace{2pt} I_n(k,\e,\s),
\ee
where $I_n(k,\e,\s)$ depends on the theory and carries information about the external particles, namely their momentum $k$ and polarization vectors $\e$.  Invariance of the integrand in \eqref{Mn=} under $SL(2,\mathbb{C})$ transformations restricts the form of $I_n(k,\e,\s)$.  Finally, the delta functions appearing in \eqref{Mn=} completely localize all integrals.

We will now present how the scattering equations are used in the case of Yang-Mills amplitudes, since later we will make use of the formulas in our examples.  From \cite{Cachazo:2013hca} we know that after performing the integration \eqref{Mn=} the tree level $n$-gluon partial amplitude  $A_n$ of Yang-Mills in arbitrary dimensions can be expressed through
\be \label{An=}
A_n = \sum_{{\rm roots}} \frac{1}{\s_{12}\s_{23}\cdots \s_{n1}} \frac{{\rm Pf}' \Psi(k,\e,\s)}{{\rm det}' \Phi},
\ee
where the sum runs over all solutions of \eqref{scattering equations}.  The $2n \times 2n$ antisymmetric matrix $\Psi$ is given by
\be 
\Psi =
\left( 
\begin{matrix}
A & -C^{\rm T} \\
C & B
\end{matrix}
\right),
\ee
where the $n\times n$ matrices $A,~B,~C$ are given by
\be
A_{ab} = \frac{k_a \sdot k_b}{\s_{ab}} \d_{a\neq b}, \quad
B_{ab} = \frac{\e_a \sdot \e_b}{\s_{ab}} \d_{a\neq b}, \quad
C_{ab} = \frac{\e_a \sdot k_b}{\s_{ab}} \d_{a\neq b}
-\d_{ab} \sum_{c\neq a}^n \frac{\e_a \sdot k_c}{\s_{ac}}.
\ee
The matrix $\Phi$ is defined as
\be 
\Phi_{ab} = \frac{\partial f_a}{\partial \s_b} = 
\frac{k_a \sdot k_b}{\s_{ab}^2} \d_{a\neq b}
-\d_{ab} \sum_{c\neq a}^n \frac{k_a \sdot k_c}{\s_{ac}^2}
\ee
and the primes in \eqref{An=} denote
\be \label{primes}
{\rm Pf}' \Psi = 2\frac{(-1)^{i+j}}{\s_{ij}} {\rm Pf} (\Psi^{ij}_{ij}),\quad
{\rm det}' \Phi = \frac{{\rm det}(\Phi^{ijk}_{pqr})}{(\s_{ij}\s_{jk}\s_{ki})(\s_{pq}\s_{qr}\s_{rp})}.
\ee
The matrix $\Psi^{ij}_{ij}$ in \eqref{primes} is derived from the matrix $\Psi$ after the removal of the $i^{\rm th}$ and $j^{\rm th}$ row and the $i^{\rm th}$ and $j^{\rm th}$ column, with $1\leq i<j\leq n$.  Finally, the matrix $\Phi^{ijk}_{pqr}$ is derived from the matrix $\Phi$ after removing the $\{i,j,k\}$ rows and the $\{p,q,r\}$ columns. 

\section{An algorithm to evaluate the amplitudes}

The first question we would like to address is whether the scattering equations can be explicitly solved and whether such a solution is useful for calculating the amplitudes.  As it was shown in \cite{Dolan:2014ega} the scattering equations admit a polynomial form.  According to \cite{Dolan:2014ega} one can use the elimination theory and completely decouple the scattering equations.  Then one ends up with a one variable $(n-3)!$ degree polynomial equation, $p(\sigma_i)=0$, for one of the $(n-3)$ variables $\sigma_i$, whereas the rest $(n-4)$ variables can be uniquely determined from the solution of the aforementioned polynomial.  The coefficients of $p(\sigma_i)$ depend only on products and sums of the kinematic invariants $k_i \sdot k_j$ and in general can be very long to write them explicitly.  For the simplest non trivial case, $n=5$, the solution of $p(\sigma_i)=0$ already occupies several lines, whereas for the next case, $n=6$, as explicit calculations for special kinematics show, the six solutions become long in an uninspiring way and extend into several pages.  

In general it is not known whether one can explicitly solve any higher case.  Fortunately, one does not have to do so in order to evaluate the amplitude.  The explanation is simple.  One can in principle obtain all amplitudes of any known theory that admits a scattering equation description using the following procedure, which does not require the explicit solution of the scattering equations.  The idea is the following.  One uses the results of \cite{Dolan:2014ega} in order to write the amplitude in terms of only one variable.  The answer for the amplitude will then be a ratio of two polynomials of the same variable of degree much higher than $(n-3)!$.  Then one can iteratively use the scattering equation of the remaining variable and bring the amplitude to the form of a ratio of two polynomials of degree $(n-3)! - 1$.  Then one can use the well known Vieta formulas that associate the sum of roots of a polynomial to its coefficients and obtain the amplitude as a rational function of the kinematic invariants. 

Let us illustrate the algorithm with the help of the following toy model.  We consider the toy scattering equation to be 
\be\label{toy_scattering_equation}
x^2 - a x + b = 0
\ee 
and the toy amplitude to be
\be\label{toy_amplitude}
A_{{\rm toy}} = \sum_{{\rm roots}} \frac{x^4 - c}{x^3 - d}.
\ee
Iterative use of \eqref{toy_scattering_equation} in \eqref{toy_amplitude} gives
\be\label{toy_amplitude2}
A_{{\rm toy}} = \sum_{{\rm roots}} \frac{(a^3-2ab)x+(b^2-a^2 b-c)}{(a^2-b)x-(ab+d)}.
\ee
Let the two solutions of \eqref{toy_scattering_equation} be denoted by $r_1$ and $r_2$.  Substitution to \eqref{toy_amplitude2} yields
\be\ba 
A_{{\rm toy}} &= \frac{(a^3-2ab)r_1+(b^2-a^2 b-c)}{(a^2-b)r_1-(ab+d)}
+ \frac{(a^3-2ab)r_2+(b^2-a^2 b-c)}{(a^2-b)r_2-(ab+d)} \\
&= \frac{c_1 +c_2 (r_1+r_2) + c_3 r_1 r_2}{c_4 +c_5 (r_1+r_2) + c_6 r_1 r_2}
= \frac{c_1 +c_2 a + c_3 b}{c_4 +c_5 a + c_6 b},
\ea\ee
where the constants $c_i$ are simple functions of $a,b,c,d$ that can be easily computed.  In the above we have made use of the Vieta formula that states that the sum of roots of \eqref{toy_scattering_equation} is $r_1+r_2 = a$ and the product $r_1 r_2 = b$.  Our toy model can be easily applied to the most general case.  One has a polynomial scattering equation of degree higher than two, therefore the final expression of the amplitude contains not only the sum and the product of roots of the scattering equation, but all the elementary symmetric polynomials of the roots.  Hence, we conclude that one does not need to solve the scattering equations, whereas at the same time we have shown that the amplitude is a rational function of the kinematic invariants $k_i \sdot k_j$ as expected.

Although the above algorithm does not require the explicit solution of the scattering equations, it becomes quickly complicated.  The authors of \cite{Dolan:2014ega} have stopped the demonstration of their construction at $n=6$.  Although in principle it is possible to extend the analysis to higher cases, it becomes difficult to continue beyond $n=6$ or $n=7$ and one should perhaps rely on other ideas in order to explicitly obtain, organize or write down the amplitude.  One such idea is the expression of the amplitudes with the help of a generating function, to which we now turn.  

\section{Calculation of the generating function}

All $n=5$ amplitudes can be decomposed as sums of the following fundamental quantity
\be \label{P1=}
P_{\vec{\a}} \equiv P  = \sum_{{\rm roots}} \frac{1}{{\rm det'} \Phi} \frac{1}{{\s_{12}}^{2+\a_1} {\s_{23}}^{2+\a_2} {\s_{34}}^{2+\a_3} {\s_{45}}^{2+\a_4} {\s_{15}}^{2+\a_5} {\s_{13}}^{\a_6} {\s_{14}}^{\a_7} {\s_{24}}^{\a_8} {\s_{25}}^{\a_9} {\s_{35}}^{\a_{10}}},
\ee
with momentum and helicity dependent coefficients.  In the above, the $\a_i$s are assumed to be integers.  Such a decomposition might not be immediately obvious and we explain how this can be done later in an example.  We demand $SL(2,\mathbb{C})$ invariance that will fix five of the $\a_i$s to the values
\be\ba \label{alpha=}
\a_6 & = -\a_1-\a_2+\a_4,\qquad
&\a_7  = +\a_2-\a_4-\a_5,\\
\a_8 & = -\a_2-\a_3+\a_5,\qquad
&\a_9  = -\a_1+\a_3-\a_5,\\
\a_{10} & = +\a_1-\a_3-\a_4.
\ea\ee
In the definition of $P$ in \eqref{P1=} one can shift any of the $\a_i$s by an integer value without loss of generality.  Since there is no canonical way to express that, we have chosen $P$ to correspond to the color ordered $\phi^3$ amplitude \cite{Cachazo:2013iea} when we set $\a_i = 0,~i=1,\ldots,5$.  Other starting points can also be considered and lead to final expressions of the same or higher complexity.

Substitution of \eqref{alpha=} in \eqref{P1=} yields
\be \label{P2=}
P = \sum_{{\rm roots}} \frac{1}{{\rm det'} \Phi}  
  \prod_{i=1}^5 \frac{1}{\s_{i,i+1}^2}
  \left(\frac{\s_{i,i+2} \s_{i+1,i+4}}{\s_{i,i+1}\s_{i+2,i+4}}\right)^{\a_i}.
\ee
The number of cross ratios appearing in \eqref{P2=} coincides with the number of independent cross ratios in $d$-dimensions, namely $n(n-3)/2$.  Since our problem is one dimensional the number of independent cross ratios in our case is $n-3$, which means that the conformal ratios in \eqref{P2=} are dependent, in general in a complicated way.  We have traded away this complication by considering a larger set of cross ratios that has the advantage that all remaining cross ratios can be simply expressed as products of quantities of that set.

The scattering equations for the $n=5$ case are quadratic in nature and admit two solutions that contain one square root.  After we fix the $SL(2,\mathbb{C})$ invariance and substitute the solution of the scattering equations in $P$ we get an expression of the following form
\be \label{P3=}
P  = \frac 1 2 (b_0+c_0 \sqrt{r}) \prod_{i=1}^5(b_i+c_i \sqrt{r})^{\a_i}
+\frac 1 2 (b_0-c_0 \sqrt{r}) \prod_{i=1}^5(b_i-c_i \sqrt{r})^{\a_i},
\ee 
where the $b_0, b_i, c_0, c_i, r$ appearing in the above expression are rational functions of the kinematic invariants $k_i \sdot k_j$ and in general can be complicated.  $b_0, c_0$ depend on how we choose to parametrize the $\a_i = 0$ case.  For integer $\a_i$s the quantity $P$ is a rational function of the kinematic invariants.  Here and in the rest of this work the index $i$ always takes the values $1,\ldots,5$ and never zero.  We consider cyclicity of indices and identify $i + 5 \sim i$.

There is a nice way in mathematics to express \eqref{P3=} via a generating function namely 
\be\label{P4=} 
P = \left. \left( \prod_{i=1}^5 \frac{1}{\a_i !} \frac{\partial^{\a_i}}{\partial x_i^{\a_i}} \right) G(x_i) \right|_{x_i = 0} .
\ee
We first consider the case where all $\a_i$s are positive. We find that the generating function $G(x_i)$ is given by
\be\label{G=}
G(x_i) = \frac{
\sum_{i<j<k<l<m}^5 (d_0 + d_i x_i + d_{ij} x_i x_j + d_{ijk} x_i x_j x_k + d_{ijkl} x_i x_j x_k x_l +d_{ijklm} x_i x_j x_k x_l x_m)
}
{\prod_{i=1}^5 (1-2 b_i x_i+(b_i^2-r c_i^2)x_i^2)},
\ee
where
\be\ba\label{d=}
d_0 & = b_0,\\
d_i & = -b_i d_0 + r c_i c_0 \equiv -b_i d_0 - r c_i f_0 ,\\
d_{ij} & = -b_j d_i + r c_j (b_i f_0+ c_i d_0) \equiv -b_j d_i - r c_j f_i,\\
d_{ijk} & = - b_k d_{ij} + r c_k (b_j f_i+ c_j d_i) \equiv  - b_k d_{ij} - r c_k f_{ij},\\
d_{ijkl} & = - b_l d_{ijk} + r c_l (b_k f_{ij} + c_k d_{ij}) \equiv - b_l d_{ijk} - r c_l f_{ijk},\\
d_{ijklm} &= -b_m d_{ijkl} + r c_m (b_l f_{ijk} + c_l d_{ijk}).
\ea\ee
We see that the $d_{ij\ldots}$ coefficients have the structure of nested sums.

So far we have considered the case $a_i > 0$.  When one or more of the $a_i$s are negative we simply make the replacement in the generating function $b_j \rightarrow b_j/(b_j^2-r c_j^2)$ and $c_j \rightarrow -c_j/(b_j^2-r c_j^2)$, for every $j$ with $a_j$ negative.

We need to express \eqref{G=} in terms of kinematic data.  This can be achieved without the need of knowing the explicit solutions of the scattering equations by considering special cases of \eqref{P3=} and then using our algorithm in order to evaluate these special cases.  There are many ways to do this and here we provide one of them.  We have 
\be\ba
b_0 &= P_{(0,0,0,0,0)}, \quad
b_1^2-r c_1^2 = \frac{2 b_0 b_1 - P_{(1,0,0,0,0)}}{P_{(-1,0,0,0,0)}} ,\\
b_1 &= \frac{ P_{(2,0,0,0,0)} P_{(-1,0,0,0,0)}-b_0 \, P_{(1,0,0,0,0)} }
             { 2 P_{(1,0,0,0,0)}P_{(-1,0,0,0,0)}-2 b_0^2},
\ea\ee
and similarly for the remaining indices.  For the numerator of \eqref{G=} we have $d_1 = P_{(1,0,0,0,0)} - 2 b_0 b_1$, whereas in order to find $d_{12}$ we can consider the $P_{(1,1,0,0,0)}$ case, in order to find $d_{123}$ the $P_{(1,1,1,0,0)}$ case and so on.  Performing the algebra we find
\be\ba\label{coefficients}
b_0 & = \sum_{i=1}^5\frac{1}{k_{i,i+1}k_{i+2,i+3}},~~
2 c_i =\pm \frac{1}{k_{i,i+1} k_{i+2,i+4}},~~ 
b_i^2 - r c_i^2  = \frac{k_{i,i+2} k_{i+1,i+4}}{k_{i,i+1} k_{i+2,i+4}},\\ 
2b_i & = \frac{k_{i+1,i+3} k_{i+2,i+3}}{k_{i,i+1} k_{i+2,i+4}} -\frac{k_{i,i+2}}{k_{i+2,i+4}}- \frac{k_{i+1,i+4}}{k_{i,i+1}} = \frac{k_{i,i+3} k_{i+3,i+4}}{k_{i,i+1} k_{i+2,i+4}} -\frac{k_{i,i+2}}{k_{i,i+1}}- \frac{k_{i+1,i+4}}{k_{i+2,i+4}},\\
-b_i b_0 + r c_i c_0 &=  \frac{k_{i,i+2}k_{i+1,i+4}(k_{i,i+3}k_{i+2,i+3}+k_{i,i+3}k_{i+1,i+2}+k_{i+1,i+2}k_{i+1,i+3})}{k_{12}k_{23}k_{34}k_{45}k_{15}k_{i+2,i+4}}.
\ea\ee 
This is enough data to determine the generating function for all of the $2^5=32$ possibilities of signs of the different $\a_i$s.  From \eqref{P3=} we see that we can simultaneously change the signs of $c_0$ and $c_i$s without altering the final result.  This can also be seen from the fact that the coefficients of the generating function always involve products of $c_0$ and $c_i$s that cancel the sign.  In \eqref{coefficients} we have given two equivalent expressions for the $b_i$s.  One can go from one to the other using conservation of momentum.  One might think that since in \eqref{d=} we have nested sums, the coefficients will become increasingly complicated.  This is not necessarily the case, since cancellations can and do occur.  For example, in the case of all $\a_i>0$ we simply have $d_{ijklm}=0$.  We should finally mention that for the $n=5$ case one can alternatively  use the explicit solutions of the scattering equations in order to find the coefficients in \eqref{coefficients}.  

For completeness, let us briefly consider the $n=4$ case.  We have that the amplitudes are linear combinations of the following fundamental quantity
\be\label{n=4}
\sum_{{\rm roots}} \frac{1}{{\rm det'} \Phi}  
  \prod_{i=1}^4 \frac{1}{\s_{i,i+1}^2}
  \left(\frac{\s_{13} \s_{24}}{\s_{12}\s_{34}}\right)^{\a_1}
  \left(\frac{\s_{13} \s_{24}}{\s_{23}\s_{14}}\right)^{\a_2}.
\ee
Since we only have one root that we need to sum over, we can explicitly evaluate \eqref{n=4} to be
\be 
-\left( \frac{1}{k_{12}}+\frac{1}{k_{23}} \right)
\left( 1+\frac{k_{23}}{k_{12}} \right)^{\a_1}
\left( 1+\frac{k_{12}}{k_{23}} \right)^{\a_2}.
\ee
One can easily find a generating function that reproduces the result.  For example, for $\a_i>0$ we get
\be \label{n=4b}
-\left( \frac{1}{k_{12}}+\frac{1}{k_{23}} \right)
\left[ 
\left( 1- \left(1+\frac{k_{23}}{k_{12}} \right) x_1 \right)
\left( 1- \left(1+\frac{k_{12}}{k_{23}} \right) x_2 \right) 
\right]^{-1}.
\ee
We see that for $x_i=0$ the generating function gives the expected answer for the color ordered $\phi^3$ theory.

\section{Examples}
As we have mentioned the amplitude is not always manifestly expressed as a linear combination of the fundamental quantity \eqref{P1=}.  Nevertheless, we can always bring it to the desired form and we demonstrate how this can be done in the case of Yang-Mills.  

We choose to remove the first two rows and columns of the reduced pfaffian in \eqref{An=}.  Then we are left with an $8\times8$ antisymmetric matrix, whose naive pfaffian expansion contains 105 terms.  Most of the terms are already of the form \eqref{P1=}, except for three cases.  The first case involves one diagonal element of the matrix $C_{ab}$, the second case involves a product of two diagonal elements of $C_{ab}$, whereas the last case consists of three diagonal elements.  In all of the three cases the treatment is the same.  We pick up one of the four terms of each $C_{aa}$ and we replace it using conservation of momentum.  Let us illustrate this with an example.  Upon expanding the reduced pfaffian of the Yang-Mills, one of the 105 terms has the form
\be 
\frac{1}{\s_{24}\s_{15}\s_{45}} 
\left(\frac{\e_3\sdot k_1}{\s_{31}} 
+\frac{\e_3\sdot k_2}{\s_{32}} 
+\frac{\e_3\sdot k_4}{\s_{34}} 
+\frac{\e_3\sdot k_5}{\s_{35}} \right).
\ee
We have omitted an overall factor that depends purely on helicities and momenta and we have kept only the part that contains the $\s_i$ variables.  We now replace in the above expression the momentum $k_1 \rightarrow -k_2-k_3-k_4-k_5$ to get
\be 
\frac{1}{\s_{24}\s_{15}\s_{45}} 
\left( 
-\frac{\s_{12}}{\s_{13}\s_{23}} \e_3\sdot k_2 
+\frac{\s_{14}}{\s_{13}\s_{34}} \e_3\sdot k_4
+\frac{\s_{15}}{\s_{13}\s_{35}} \e_3\sdot k_5
\right).
\ee
We have now achieved our goal.  Combining all elements together the contribution of our term to the scattering amplitude is proportional to
\be
\sum_{{\rm roots}} \frac{1}{{\rm det'} \Phi}  
  \prod_{i=1}^5 \frac{1}{\s_{i,i+1}^2}
  \left(
     \frac{\s_{12}\s_{34}}{\s_{13}\s_{24}} \e_3\sdot k_2 
     -\frac{\s_{14}\s_{23}}{\s_{13}\s_{24}} \e_3\sdot k_4 
     -\frac{\s_{15}\s_{23}\s_{34}}{\s_{13}\s_{24}\s_{35}} \e_3\sdot k_5 
  \right).
\ee
In terms of our fundamental quantity \eqref{P1=}, the above expression becomes
\be \label{example}
      \e_3\sdot k_2 P_{(-1,0,-1,0,0)} 
     -\e_3\sdot k_4 P_{(0,-1,0,0,0)}
     -\e_3\sdot k_5 P_{(0,-1,-1,0,-1)}.
\ee
The three $P$s appearing in \eqref{example} can be easily computed from \eqref{P4=}.  We find
\be
P_{(0,-1,0,0,0)} = \frac{b_2 d_0-r c_2 c_0}{b_2^2-r c_2^2} = \frac{1}{k_{12}k_{45}}+\frac{1}{k_{12}k_{34}}+\frac{1}{k_{15}k_{34}},
\ee
where the denominator $b_2^2-r c_2^2$ comes from the fact that the $\a_2$ is negative in our example.  The same way we find
\be 
P_{(-1,0,-1,0,0)} = \frac{(b_1 b_3+r c_1 c_3)d_0-(b_1 c_3+b_3 c_1)r c_0}{(b_1^2-r c_1^2)(b_3^2-r c_3^2)}
=\frac{1}{k_{15}k_{23}}+\frac{1}{k_{23}k_{45}},
\ee
and finally. 
\be 
P_{(0,-1,-1,0,-1)} = \frac{1}{k_{12}k_{45}}.
\ee

The case of the full Yang-Mills and gravity can be treated the same way, whereas for more general theories we expect similar considerations.

\section{Discussion}

The main motivation of this work was to nicely organize and even calculate the tree amplitudes of theories whose S-matrix can be described through the scattering equations.  In doing so, we considered the most general quantity consistent with $SL(2,\mathbb{C})$ invariance, that in general depends on conformal cross ratios of the variables that appear in the scattering equation.  Then, all amplitudes can be written as linear combinations of this quantity, with coefficients that depend on kinematic data.  We have found that our fundamental quantity can be nicely expressed through a generating function, that we have explicitly calculated for the first non-trivial case, namely $n=5$.  Although the solutions of the scattering equations are complicated in nature, we have argued that knowledge of them is not necessary to evaluate the amplitude.  We have also presented a simple argument why the amplitude is a rational function of the kinematic invariants.

Although all five point amplitudes can also be obtained by brute force, it might be time consuming to simplify all square roots appearing, whereas the final answer can be given in a disorganized form.  Experience shows that for practical purposes the computation can be simplified using the polynomial form of the scattering equations, but even that does not solve the problem of organization.  Our proposed generating function gives the answer in a neat way.  One can also see that it is rational in the kinematic invariants, it has the structure of nested sums and it knows about all signs coming from combinatorics.

Our method can certainly be applied to the $n=6$ case using our algorithm and the results of \cite{Dolan:2014ega}, whereas a generalization to the arbitrary $n$ case remains to be seen.  The general case is expected to be captured by a generating function of $n(n-3)/2$ variables, which is simply the number of all possible $\s_{ij}$s with $i<j$ after we subtract the $n$ conditions coming from the $SL(2,\mathbb{C})$ invariance of the problem.  The form of our fundamental quantity \eqref{P1=} can be easily extended to the general case and it involves the product of $n(n-3)/2$ cross ratios each one appearing an integer number of times.  The form of the generating function in the general case is also easy to be found.  It will involve a denominator of $n(n-3)/2$ polynomials, each one of degree $(n-3)!$.  The difficult part of the computation is to express the various coefficients of the generating function in terms of kinematic data.  One possible way to achieve this would be to assume the form of the generating function and fix its coefficients by studying simple cases, where we already know the answer.  It is also worth to investigate whether there is an even simpler but equivalent form of our generating function for $n=5$, having always in our mind application to the general case.

\vspace{3mm}

\noindent
{\bf Acknowledgments}

\vspace{3mm}

\noindent
It is a pleasure to thank Nima Arkani-Hamed, Wei He, Gregory Korchemsky and Francisco Rojas for useful comments and discussions.  We thank the organizers of the \textit{`3rd Joint Dutch-Brazil School on Theoretical Physics'} and the organizers of \textit{`Program on Integrability, Holography and the Conformal Bootstrap'}, that took place at the ICTP-SAIFR in S\~{a}o Paulo, for creating an inspiring environment.  The work of C.K. is supported by the S\~ao Paulo Research Foundation (FAPESP) under grants 2011/11973-4 and 2012/00756-5.

\bibliographystyle{utphys}
\bibliography{mybib}
\end{document}